# Intrinsic Electrical Transport and Performance Projections of Synthetic Monolayer MoS₂ Devices

*Kirby K. H. Smithe, Chris D. English, Saurabh V. Suryavanshi, and Eric Pop*

Department of Electrical Engineering, Stanford University, Stanford, CA 94305, U.S.A.

**Abstract:** We demonstrate monolayer MoS₂ grown by chemical vapor deposition (CVD) with transport properties comparable to those of the best exfoliated devices over a wide range of carrier densities (up to ~$10^{13}$ cm$^{-2}$) and temperatures (80-500 K). Transfer length measurements (TLM) decouple the intrinsic material mobility from the contact resistance, at practical carrier densities (>$10^{12}$ cm$^{-2}$). We demonstrate the highest current density reported to date (~270 µA/µm or 44 MA/cm²) at 300 K for an 80 nm device from CVD-grown monolayer MoS₂. Using simulations, we discuss what improvements of monolayer MoS₂ are still required to meet technology roadmap requirements for low power (LP) and high performance (HP) applications. Such results are an important step towards large-area electronics based on monolayer semiconductors.

**E-mail:** epop@stanford.edu

## 1. Introduction

Monolayer (1L) two-dimensional (2D) semiconductors such as $MoS_2$ have garnered attention for highly scaled optoelectronics and flexible electronics due to their sub-nm thickness, direct band gap, and lack of dangling bonds [1,2]. For practical applications, such films must be grown over large areas and must demonstrate good electrical properties. Until recently [3–9], however, the highest reported mobility of 1L $MoS_2$ field effect transistors (FETs) grown by chemical vapor deposition (CVD) had been below 20 cm²/V/s on isolated single crystals [10–24]. In addition, little systematic work has been done to understand metallic contacts to CVD-grown 1L films [25], which ultimately limit device performance with scaling.

Here we present the first rigorous transfer length method (TLM) study of as-grown 1L CVD $MoS_2$ devices as a function of temperature to systematically separate contributions to total device resistance ($R_{TOT}$) from contacts and the channel. Building on previous work to improve contact resistance ($R_C$) [26] we obtain $R_C \approx 6.5$ k$\Omega$·µm at room temperature and moderate carrier densities. We also extract the effective electron mobility ($\mu_{eff}$) from sheet resistance measurements to be 20 cm²/V/s, comparable to unencapsulated exfoliated 1L devices on $SiO_2$. Fitting with our compact model [27,28] yields similar values for $R_C$ and $\mu_{eff}$ and allows us to simulate aggressively scaled device channel lengths ($L$) while maintaining the key material properties. With an equivalent oxide thickness (EOT) and values for $R_C$ as dictated by future ITRS (International Technology Roadmap for Semiconductors) nodes, the simulations predict that for both the high performance (HP) and low power (LP) specifications, the maximum achievable on-state current ($I_{ON}$) is more strongly dependent on the saturation velocity, $v_{sat}$, than on mobility.

## 2. Methods

We synthesize continuous 1L $MoS_2$ from solid S and $MoO_3$ precursors with the aid of perylene-3,4,9,10 tetracarboxylic acid tetrapotassium salt (PTAS) [11,29–31] on $SiO_2$ on Si (p++) substrates, which also serve as back-gates for field-effect devices. Elevated temperature (850 °C) and atmospheric pressure are utilized to encourage lateral epitaxial growth (Figures S1 and S2), and the CVD conditions can be tailored to produce either a continuous 1L film or single-crystal domains up to $10^5$ µm² (triangular crystals with edges exceeding 300 µm, see Figure 1a-d and Supplementary Figure S3). Key advances in this work include a combination of PTAS seeding around the chip perimeter, higher synthesis temperatures, and improved electrical contacts with pure Au, which lead to the improved device results shown here. Previous studies that implemented PTAS for CVD growth did not have high-quality electrical data due to (relatively) poor electrical contacts or small grain sizes. Conversely, previous studies that made contact improvements focused on exfoliated multi-layer $MoS_2$, not on CVD-grown monolayers. Additional discussion about various growth conditions and their optimization is provided in the Supplement.

We define rectangular channel regions by $XeF_2$ etching, and TLM [32] structures with varying channel lengths ($L = 80$ nm to 1.2 μm) by electron beam (e-beam) lithography, exclusively on 1L regions, as shown in Figures 1e and 1f. Pure Au contacts deposited by e-beam evaporation under high vacuum ($\sim 5 \times 10^{-8}$ Torr) are employed without any adhesion layer to achieve a clean contact interface and reduce contact resistance [26]. All electrical measurements were carried out in a vacuum probe station ($\sim 10^{-5}$ Torr) following a vacuum anneal *in situ* at 200 °C for 1 hour.

Atomic force microscopy (AFM), Raman spectroscopy, and photoluminescence (PL) are utilized post-fabrication to confirm that the $MoS_2$ devices are indeed 1L [33–35], and that their optical and excitonic properties have been preserved. Figure 2a illustrates an AFM step-height profile of ~1 nm, which is consistent with the 1L $MoS_2$ thickness plus a van der Waals gap. Figure 2b depicts the in-plane and out-of-plane Raman modes, which are sometimes incorrectly labeled $E_{2g}$ and $A_{1g}$ in the literature; however, this notation is only strictly correct for bulk and even-number-layer samples, which belong to the $D_{6h}$ and $D_{3d}$ point groups, respectively. Odd-numbered few-layer $MoS_2$ samples (including 1L) belong to the $D_{3h}$ point group. Thus, these Raman features are denoted as E' and $A_1'$ [36–38] at 384.3 and 403.7 cm$^{-1}$, with a peak separation $\Delta f \sim 20$ cm$^{-1}$ [4,5,9] typical for as-grown $MoS_2$ monolayers with slight intrinsic tensile strain. Lastly, the A and B exciton peaks, estimated to be separated by a valence band splitting of ~150 meV at the K point [35,39] are clearly exhibited with peaks at 1.79 and 1.93 eV.

## 3. Experimental Results and Discussion

Typical transfer curves for these devices with varying channel lengths are shown in Figure 3a at a drain bias $V_{DS} = 1$ V. The carrier density ($n$) is estimated by assuming a simple linear charge dependence on the gate voltage overdrive

$$n \approx \frac{C_{ox}}{q}\left(V_{GS} - V_T\right) \qquad\qquad (1)$$

where $C_{ox} \approx 38$ nF/cm$^2$ is the capacitance per unit area of $t_{ox} = 90$ nm SiO$_2$, $V_{GS}$ is the gate-source voltage, and $V_T$ is the threshold voltage obtained by the linear extrapolation method [32] as demonstrated in Figure 3a (dashed lines fit to the curve at maximum transconductance). Supplementary Figure S4 displays log-scale and forward-backward $I_D - V_{GS}$ sweeps of the same devices, demonstrating $I_{ON}/I_{OFF}$ of at least $10^4$ and minimal hysteresis.

Figure 3b demonstrates good least-squares fitting to $R_{TOT}$ vs. $L$ for various calculated values of $n$, suggesting uniform material and contacts for our devices. The slopes of these lines correspond to sheet resistance $R_{SH}$ [kΩ/□], whereas the ordinate intercept yields twice the width-normalized contact resistance, $2R_C \cdot W$ [kΩ·µm]; these quantities are extracted for multiple values of $n$. Resistance and mobility values are then carefully extracted for the same value of $n$ (i.e. the same gate overdrive, $V_{GS} - V_T$) rather than the same $V_{GS}$.

In Figure 4a, we observe that $R_C$ varies with $n$ as the back gate modulates the Fermi level underneath the contacts as well as in the channel. We extract a lowest $R_C = 6.5 \pm 1.5$ kΩ·µm at 300 K for $n \approx 4 \times 10^{12}$ cm$^{-2}$, with the uncertainty reflecting 90% confidence intervals from a least squares fit of the TLM curve. Although lower $R_C$ has been achieved in multilayer exfoliated MoS$_2$ FETs [26,40], this value for 1L CVD MoS$_2$ could potentially be reduced further with the aid of chemical doping techniques [40–42] or phase engineering [25,43]. Moreover, fitting with our compact model [27,28] and extrapolating to a higher $n = 10^{13}$ cm$^{-2}$, $R_C$ could drop to 5.5 kΩ·µm, without any kind of molecular doping or threshold shifting, for pure Au contacts. $R_C$ also decreases with increasing temperature ($T$), consistent with increased thermionic emission over the Schottky barrier at the contacts. Lastly, we note that although our devices exhibit a linear $I_D$-$V_{DS}$ relationship for low drain biases (see Supplementary Figure S5), this does not justify use of the word "Ohmic" to characterize our contacts from a band structure perspective [44]. Linear $I_D$-$V_{DS}$ curves can still be obtained at low bias across a Schottky-barrier FET as a result of carriers tunneling through the barrier, resulting in non-negligible contact resistance [45–47].

Using the transmission line model [32,48], we estimate the specific contact resistivity ($\rho_C$) as shown in Figure 4b from

$$R_C W = \frac{\rho_C}{L_T} \coth\left(\frac{L_C}{L_T}\right) \approx \sqrt{\rho_C R_{SH}} \qquad (2)$$

where $L_C = 1$ μm is the length of the contacts and $L_T$ is the current transfer length, i.e. the distance over which the current flowing in the $MoS_2$ drops to $1/e$ times the value injected at the contact edge. At 300 K, we extract $\rho_C \approx 10^{-5}$ Ω·cm², a value ~12 times higher than our best results for few-layer exfoliated $MoS_2$ at the same carrier density, $n = 4 \times 10^{12}$ cm⁻² [26]. As with $R_C$, we observe $\rho_C$ to decrease with increasing $T$ and $V_{DS}$ in Figure 4b due to enhanced thermionic and field emission, respectively. Interestingly, we do not see appreciable variation with $n$, as one might expect from a thinning of the Schottky barrier at the Au/$MoS_2$ interface, which would allow for enhanced field emission. Taken together, these two observations indicate that, while both thermionic and field emission play a role, the former is by far the dominant mechanism. This disparity should be exacerbated at lower temperatures, as fewer carriers are able to thermionically surmount the barrier, and the ratio $\rho_C(V_{DS} = 0.1$ V$) / \rho_C(V_{DS} = 1.0$ V$)$ being 3 times larger at 80 K than at 300 K supports this conclusion. The weak yet observable variation of $\rho_C$ with $n$ for $T = 80$ K and $V_{DS} = 0.1$ V, where few carriers can make it over the barrier and field emission is suppressed, further supports this notion, as an increase in tunneling can then be more easily noticed. This is in contrast to our exfoliated few-layer devices [26], which have a shorter $L_T$ (~40 nm) and $\rho_C$ that clearly decreases with $n$.

We note that the approximation in Equation (2) is only valid when $L_T \ll L_C$, in which case $\coth(L_C/L_T) \approx 1$, and we can rearrange to give

$$L_T \approx \sqrt{\frac{\rho_C}{R_{SH}}} \qquad (3)$$

whereby we extract $L_T \approx 100$ nm $\ll L_C$ at 300 K (shown in Fig. 3c), justifying our use of the approximation. As previously mentioned, both $R_C$ and $\rho_C$ decrease for higher $V_{DS}$ due to increased field emission, especially at lower temperatures where thermionic emission is suppressed. This leads to $L_T$ that is essentially constant with respect to $n$, but also decreases with increasing $T$ and $V_{DS}$ (shown in Supplementary Figure S6). This result suggests that contacts to 1L $MoS_2$ can be scaled to lengths as small as 100 nm before current crowding causes an increase in $R_C$.

Utilizing $R_{SH}$ as given by the TLM fit slopes, the effective mobility can be calculated by

$$\mu_{eff} = \left(qnR_{SH}\right)^{-1} \qquad (4)$$

where $q$ is the elementary charge and $n$ is given by Equation 1. As $n$ increases and Equation 1 better approximates the true charge in the channel, Equation 4 approaches a constant lowest value, which we take to be the "true" value for $\mu_{eff}$ in the technologically relevant high carrier density regime (see Supplementary Figure S5b).

We focus on *effective* mobility rather than field-effect mobility [$\mu_{FE} = L(\partial I_D / \partial V_{GS}) / (WC_{ox} \cdot V_{DS})$] in this work because $\mu_{eff}$ is strictly a channel material parameter that is valid for all moderate values of $n$, as shown in Figure 4c. In contrast, $\mu_{FE}$ is heavily dependent on $V_{DS}$ and $V_{GS}$ (due to the Schottky contacts), leading to a concept often referred to as a "peak mobility" as a function of gate voltage [40,49–51], and potentially resulting in under- or overestimations of the true channel mobility (see Supplementary Figures S7 and S8). The TLM analysis not only allows us to separate contributions by $R_C$ and $R_{SH}$ to $R_{TOT}$, but also allows for the direct use of $\mu_{eff}$ as a more reliable channel parameter in device models. In other words, the effective mobility is the more important figure of merit (rather than the field-effect mobility), linking materials, devices (via models, as done below), and system applications of TMDs. We also note that the $\mu_{eff}$ extracted here is a lower bound on the true band mobility [23] due to fast traps and impurities at the oxide interface, as opposed to mobility values that could be obtained from Hall measurements or from devices on a smooth, clean surface such as hexagonal boron nitride (h-BN). [52]

Analyzing extracted data for $\mu_{eff}$ vs. $T$ in Figure 4d reveals $\mu_{eff}$ to be approximately constant near ~28 cm$^2$/V/s at low $T$. The lack of variation in $\mu_{eff}$ between 100 and 200 K suggests that these values are limited by impurity scattering [53–55], possibly from particles or adsorbates deposited during device fabrication. Above 200 K, $\mu_{eff}$ is principally limited by optical phonon scattering and rolls off as ~$T^{-\gamma}$ (where $\gamma \approx 1$), falling to a value of $20 \pm 3$ cm$^2$/V/s at 300 K, comparable with the best CVD 1L devices reported so far [3–5,19–21]. (Also see Supplementary Figure S9.) A larger temperature coefficient of mobility, $\gamma$, would be indicative of stronger (intrinsic) phonon scattering. As with our extractions of $R_C$, the uncertainty reflects 90% confidence intervals.

Finally, we wish to understand how the mobility and contact resistance rigorously studied thus far manifest themselves in very small devices. To this end, we fabricated a short-channel ($L$ ~ 80 nm) FET on our monolayer CVD-grown MoS$_2$ films, and we recorded the transfer characteristics shown in Figure 5a. At room temperature, we measure the highest current density reported to date (~270 μA/μm or 44 MA/cm$^2$, taking into account the appropriate 0.615 nm monolayer thickness [37,56]) for CVD-grown 1L MoS$_2$ FETs, as shown in Figure 5b. (An overview of other measurements for reported current density and mobility in 1L CVD MoS$_2$ is provided in Supplementary Figure S9.) This is an important metric, because the intrinsic delay of a transistor is $\propto C_{ox}V_{DD}/I_{ON}$, where $V_{DD}$ is the operating voltage. In other words, it is not the intrinsic mobility of the devices that affects the circuit delay, but the total drivable current, which might ultimately be contact-limited. Nonetheless, even at high carrier density ($n$ ~ 10$^{13}$ cm$^{-2}$) we note sublinear (i.e. Schottky-like) measured $I_D$–$V_{DS}$ up to $V_{DS}$ ~ 1 V in Figure 5b, further demonstrating the need for reducing contact resistance for very short channel lengths. Values for $\mu_{eff}$, $R_C$, and maximum $I_D$ could also be further improved by suppressing the detrimental effects of the underlying oxide, i.e. by fabricating devices on h-BN, on oxides with higher phonon energies, or selecting dielectrics to sufficiently screen charged impurities [55].

## 4. Simulation Projections

Before concluding, we use simulations seeking to project how such short channel 1L MoS$_2$ FETs might behave with more idealized properties, i.e. with lower $R_C$ and properly scaled insulators. To this end, we employ our physics-based device model (described elsewhere [27] and available online [28]) and first fit it against the measured data in Figure 5b, with $\mu_{eff}$ = 20–22 cm$^2$/V/s and $R_C$ = 6–8 kΩ·μm, in good agreement with our TLM extractions described earlier. The model reveals that approximately two thirds of the applied $V_{DS}$ is dropped at the contacts of the 80 nm device, further highlighting the need for contact engineering in such small devices. At higher drain voltage in Figure 5b, the model $I_D$ begins to saturate not due to channel pinch-off, but rather due to carrier velocity saturation, which for these simulations was fit to a value of $v_{sat}$ = 7×10$^5$ cm/s.

We then use our calibrated model [28] to predict the performance of such monolayer semiconductors at scaled ITRS [57] nodes for high performance (HP) and low power (LP) applications, in Figure 6. We assume the *intrinsic* channel mobility extracted from experiments in this work and Ref. [58], but take the contact resistance ($R_C$ = 150-200 $\Omega \cdot \mu m$) as required by the ITRS for HP and LP applications. To calculate the ON current we first iteratively adjust the flatband voltage through the gate workfunction to achieve the required OFF current, $I_{OFF}$ = 100 nA/$\mu m$ for HP and 10 pA/$\mu m$ for LP. We neglect the gate leakage current and the source to drain leakage which allows us to benchmark the maximum possible performance of the MoS$_2$ devices. We note that even the highly scaled devices considered in Figure 6 are fully in the diffusive transport regime (i.e. ballistic effects do not play a major role), because the electron mean free path in MoS$_2$ is just 1 to 3 nm (see Figure S10) for mobility values between 20 and 80 cm$^2$/V/s.

In Figure 6 analysis we consider two values for velocity saturation and two values for 1L MoS$_2$ mobility, as shown. The present state-of-the-art 1L MoS$_2$ (red curves, this work on CVD MoS$_2$ and that of Ref. [58] on exfoliated MoS$_2$) with $v_{sat}$ = 10$^6$ cm/s fall short of the ITRS requirements for both HP and LP. However, this exercise highlights that the role of mobility is secondary, because at high lateral field $I_{ON}$ is more strongly limited by $v_{sat}$. On the other hand, if the saturation velocity is increased to simulated projections [59,60] of $v_{sat}$ = 3.2×10$^6$ cm/s, we find that both the HP and the LP ITRS requirements could potentially be achieved using 1L MoS$_2$ (green curves) for the shorter channel devices (< 20 nm), in agreement with recent quantum transport simulations [61].

It is relevant to inquire why ITRS specifications could be met with a $v_{sat}$ that even optimistically remains lower than that of silicon. As illustrated in Supplementary Figure S11, the carrier confinement is different in monolayer 2D materials than in typical semiconductors, such that the EOT could be smaller for 2D materials ($t_{ox,2D} < t_{ox,Si}$) even for the same physical oxide thickness. This allows monolayer 2D materials to achieve higher carrier densities for similar oxide thickness and similar overdrive voltage, and thus higher $I_{ON}$ even at lower carrier velocities, while meeting $I_{OFF}$ requirements owing to their larger band gap. Consequently, future work on monolayer 2D transistors should focus not only on improving the contact resistance (as highlighted earlier), but also on understanding and improving the charge carrier drift velocities.

## 4. Conclusions

In summary, this work presents an in-depth analysis of CVD-grown 1L MoS$_2$ FETs with material characteristics comparable to those of exfoliated devices, achieving the highest current density reported to date (~270 µA/µm or 44 MA/cm²). We show that the TLM approach provides rigorous estimates of both mobility and contact resistance, obtaining $R_C \approx 6.5$ kΩ·µm and $\rho_C \approx 10^{-5}$ Ω·cm² at room temperature and moderate carrier densities, which could be reduced further with contact engineering. We use simulations to match our experimental results and to provide insights into how 1L MoS₂ devices could behave when properly scaled down. It is revealed that $v_{sat}$ plays a greater role than $\mu_{eff}$ in determining $I_{ON}$ for aggressively scaled devices. Simulations also reveal that 1L MoS₂ could nearly meet ITRS requirements at sub-20 nm channel lengths, but further advancements in contact, dielectric, and carrier velocity engineering are still needed.

**Supporting Information**.

Experimental setup of reported and optimized growth conditions, forward and backward $I_D$-$V_{DS}$ sweeps showing minimal hysteresis, low-field $I_D$–$V_{DS}$ sweeps showing linear behavior, plots of $L_T$ vs. $n$ for various $T$ and $V_{DS}$, extractions of field-effect mobility, simulations of extracted mobility values, summary plot of reported current density vs. reported mobility in 1L CVD MoS₂, mean free path vs. mobility, and illustrations of quantum confinement effects in 2D and bulk materials.

**Author Contributions**

The manuscript was written through contributions of all authors. All authors have given approval to the final version of the manuscript.

**Notes**

The authors declare no competing financial interests.

**Acknowledgments**

Work was performed at the Stanford Nanofabrication Facility (SNF) and Stanford Nano Shared Facilities (SNSF). We acknowledge technical assistance from Dr. James McVittie for CVD system maintenance and modification, and Dr. Ted Kamins for insight and discussion about CVD processes. This work was supported in part by the Air Force Office of Scientific Research (AFOSR) grant FA9550-14-1-0251, in part by the National Science Foundation (NSF) EFRI 2-DARE grant 1542883, the NCN-NEEDS program, which is funded by the NSF contract 1227020-EEC and by the Semiconductor Research Corporation (SRC), in part by the Systems on Nanoscale  Information fabriCs (SONIC, one of six SRC STARnet Centers sponsored by MARCO and DARPA), and in part by the Stanford SystemX Alliance. K.K.H.S. and C.D.E. acknowledge partial support from the Stanford Graduate Fellowship (SGF) program and NSF Graduate Research Fellowship under Grant No. DGE-114747.

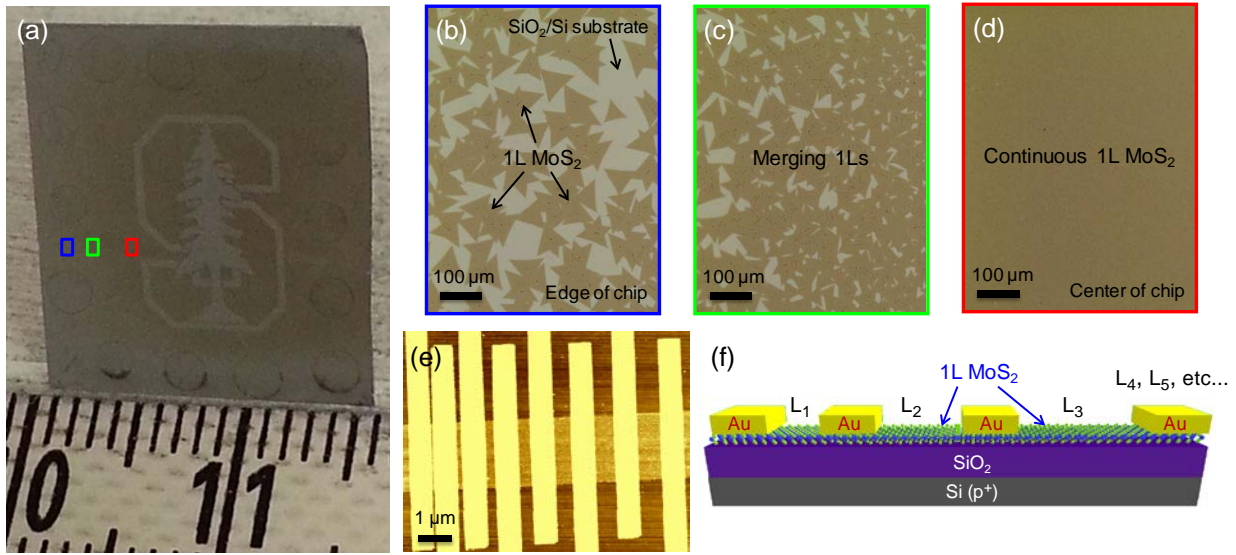

**Figure 1.** (a) Photograph of CVD MoS$_2$ nanofabric on 30 nm SiO$_2$ on Si. The Stanford logo was etched out of the continuous region for added contrast. Ruler divisions are in mm, and colored rectangles correspond to Figures 1(b-d). Circular "coffee rings" are the edges of PTAS droplets (see Supplement Section A). (b-d) Optical images of various locations on the same SiO$_2$/Si chip, taken from the edge to the center of the chip. Large triangular crystals are seen near the edge, which merge into a continuous film towards the center. Small bilayer regions (~500 nm in size) can appear depending on the growth conditions, as discussed in Supplementary Section A. (e) AFM image of TLM test structure with varying channel lengths ($L$ = 200 to 1200 nm) fabricated on 1L MoS$_2$. (f) Side-view schematic of our TLM devices with pure Au source and drain contacts.

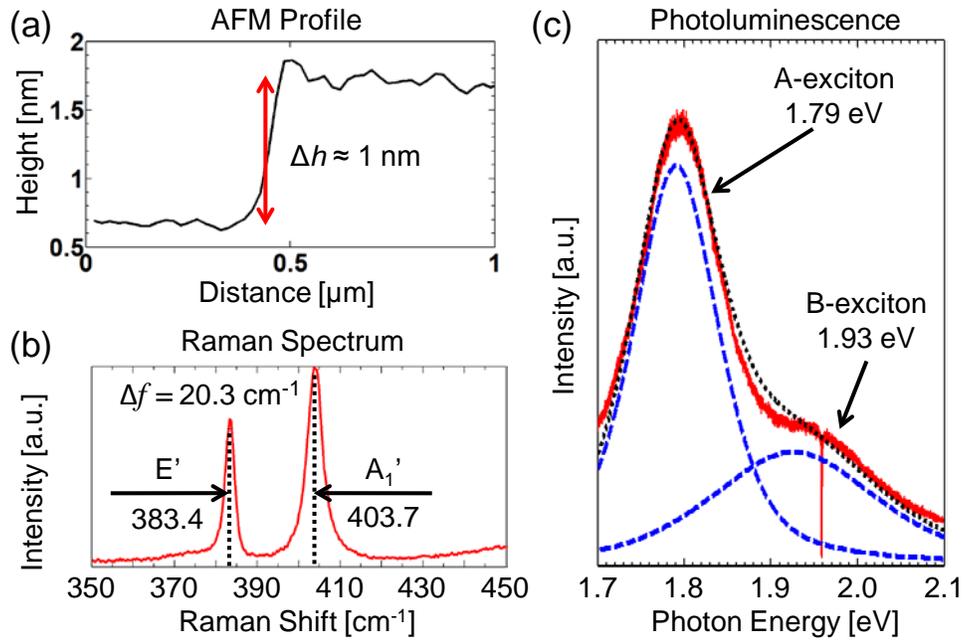

**Figure 2.** (a) AFM step height profile of a typical 1L region showing a height of ~1 nm above the 90 nm SiO$_2$ substrate. (b) Raman spectrum of our $L$ = 1.2 μm device after fabrication and measurement showing the E' and A$_1$' modes of 1L MoS$_2$. (c) Photoluminescence (red data) and Lorentzian curve fittings (blue and black dashed lines) of the same device. An excitation wavelength of 532 nm was used for all Raman and PL spectra.

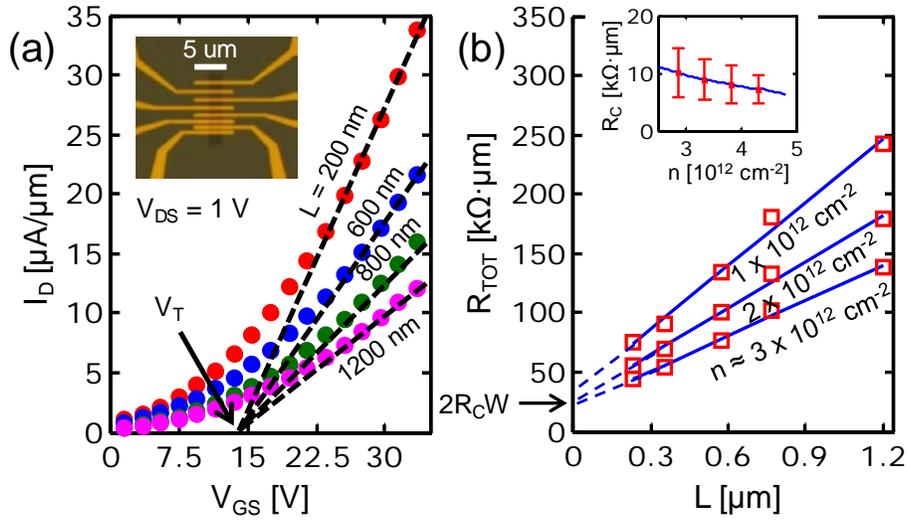

**Figure 3.** (a) Width-normalized transfer curves of 1L CVD-grown MoS$_2$ for varying channel lengths $L$ as noted. Inset: optical image of TLM structure as fabricated by e-beam lithography. The arrow points to the linearly extrapolated threshold voltage, $V_T$ (b) Plot of total device resistance $R_{TOT}$ (normalized by width) vs. $L$, yielding the sheet resistance $R_{SH}$ and contact resistance $R_C$ from the slopes and ordinate intercepts, respectively. Inset: $R_C$ vs. carrier density $n$, shown with error bars reflecting 90% least squares fit confidence intervals. Symbols are experimental data and lines are linear fits for all figures shown.

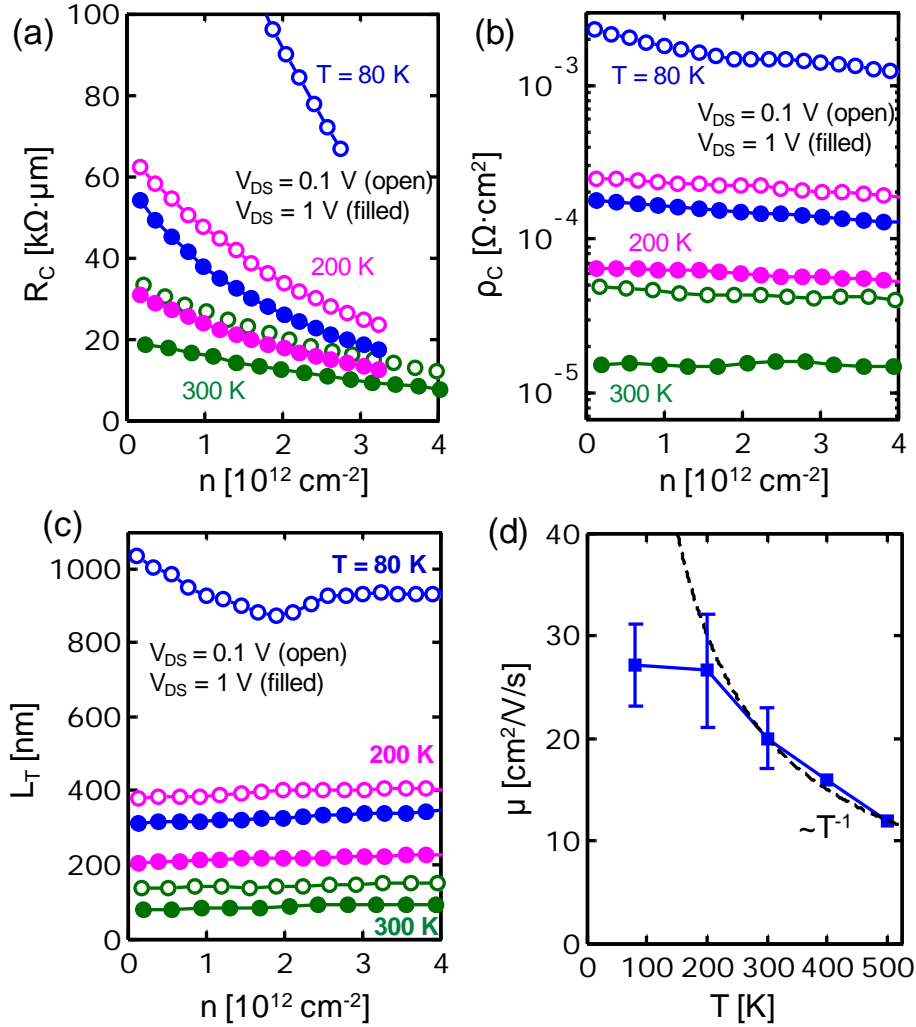

**Figure 4.** Temperature and carrier density dependence of contact resistance and mobility. (a) $R_C$ vs. $n$ at drain biases of $V_{DS} = 0.1$ (open symbols) and 1.0 V (filled) for $T = 80$, 200, and 300 K. Reduced $R_C$ with higher $T$ and higher $V_D$ is consistent with increased thermionic and field emission over and through the Schottky barrier, respectively. (b) Corresponding $\rho_C$ vs. $n$ using the transmission line model, yielding $L_T \approx 100$ nm at 300 K (also see Supplementary Figure S5.) (c) Transfer length vs. carrier density. With the exception of $V_{DS} = 0.1$ and $T = 80$ K, where both thermionic and field emission are greatly suppressed, we observe virtually no change in $L_T$ with increasing $n$, similar to our extraction of $\rho_C$. (d) $\mu_{eff}$ vs. $T$ for $n = 3 \times 10^{12}$ cm$^{-2}$. We observe $\mu_{eff}$ to be constant below ~200 K and roll off as $T^{-1}$ above. Error bars reflect 90% confidence intervals.

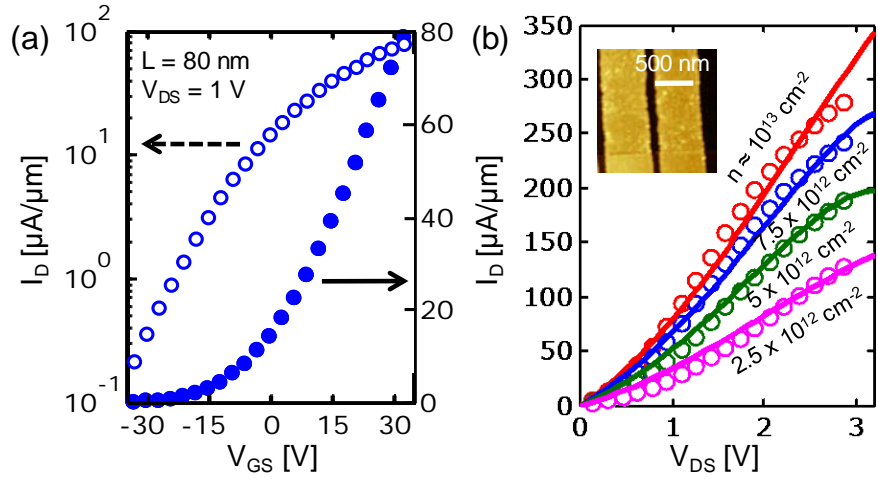

**Figure 5.** (a) Transfer curve of an $L \approx 80$ nm 1L MoS$_2$ FET. $I_{ON}/I_{OFF}$ is only ~300 for this device due to short-channel effects (back-gated with relatively thick $t_{ox} = 90$ nm). Left axis: log scale; right axis: linear scale ($W = 3.4$ μm). (b) $I_D$–$V_{DS}$ curves for the same device at varying average carrier density $n$. Symbols are measured data, demonstrating $I_D > 270$ μA/μm for $n \approx 10^{13}$ cm$^{-2}$; lines correspond to a fit with our semi-classical model. The model fits with $R_C \approx 7$ kΩ·μm and $\mu_{eff} \approx 20$ cm$^2$/V/s, consistent with our TLM extractions. Inset: AFM image of the device.

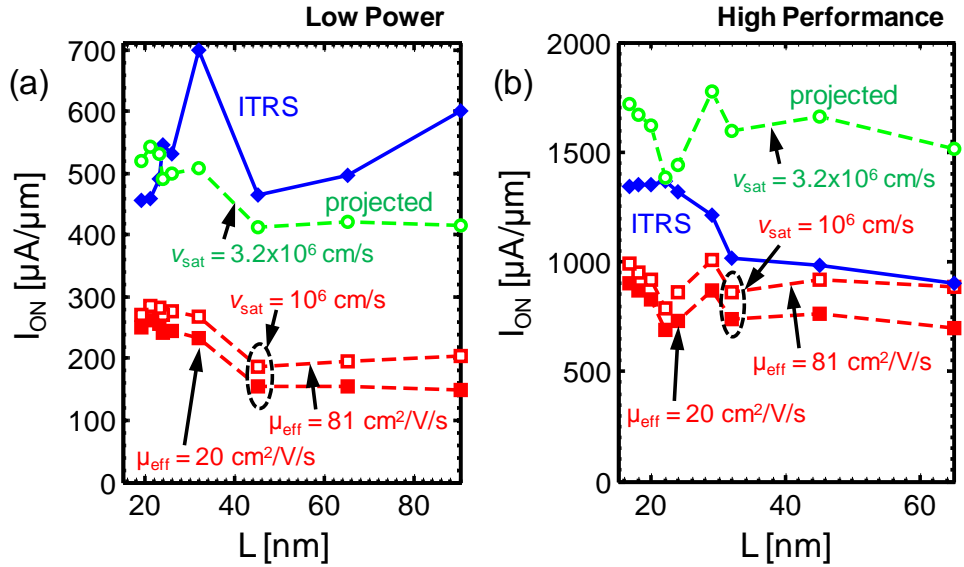

**Figure 6.** Benchmarking 1L MoS$_2$ devices for (a) low power (LP) and (b) high performance (HP) ITRS [57] requirements, listed vs. the gate length. ITRS requirements are shown in blue with fixed $I_{OFF}$ = 10 pA/μm for LP and 100 nA/μm for HP. Simulations in red use $v_{sat}$ = 10$^6$ cm/s, with solid symbols for our CVD-grown MoS$_2$ ($\mu_{eff}$ = 20 cm$^2$/V/s) and open symbols from recent exfoliated data [58] ($\mu_{eff}$ = 81 cm$^2$/V/s). The green curve shows projections that meet ITRS requirements with the higher mobility and with $v_{sat}$ = 3.2×10$^6$ cm/s, not yet demonstrated experimentally.

# Intrinsic Electrical Transport and Performance Projections of Synthetic Monolayer MoS₂ Devices

*Kirby K. H. Smithe, Chris D. English, Saurabh V. Suryavanshi, and Eric Pop*

Department of Electrical Engineering, Stanford University, Stanford, CA 94305, U.S.A.

## A. Growth Details

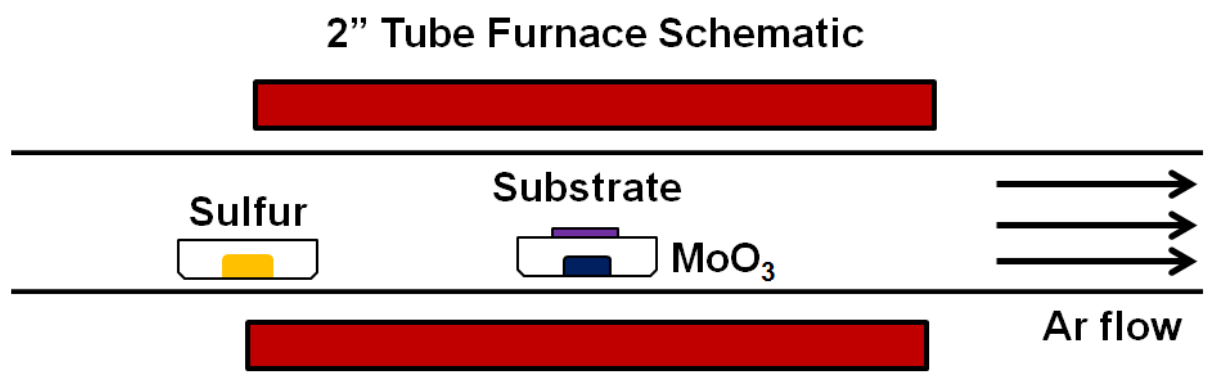

**Figure S1.** Diagram of experimental setup for large-grain 1L MoS₂ synthesis.

All growths are performed in a 2-inch inner diameter quartz tube, using a 2-inch Across International STF 1100 Tube Furnace connected to an Ebara A07 vacuum pump.

For the electrical data shown in this paper, experimental details are as follows. ~200 mg solid S was placed upstream in a quartz boat, and ~1 mg $MoO_3$ was placed in an alumina crucible liner at the center of the furnace, with 11 inches separating the precursors. A 90 nm $SiO_2$/Si substrate was treated with piranha solution (3:1 $H_2SO_4$:$H_2O_2$) for several hours before using a pipette to drop a single drop (~25 μL) of 80 μM perylene-3,4,9,10 tetracarboxylic acid tetrapotassium salt (PTAS) onto the substrate center. After drying the PTAS on a hot plate in air, the substrate was placed face-down on the crucible over the $MoO_3$. The growth recipe was:

1) Pump the tube to base pressure (~100 mTorr)

2) Ramp to 300 °C in 10 minutes while flowing 500 sccm Ar to reach 760 Torr

3) Anneal at 300 °C, 760 Torr for 5 minutes

4) Reduce the Ar flow to 10 sccm, ramp to 850 °C in 15 minutes and hold for 15 minutes

5) Allow the furnace to cool to <600 °C before opening the hatch for rapid cooling

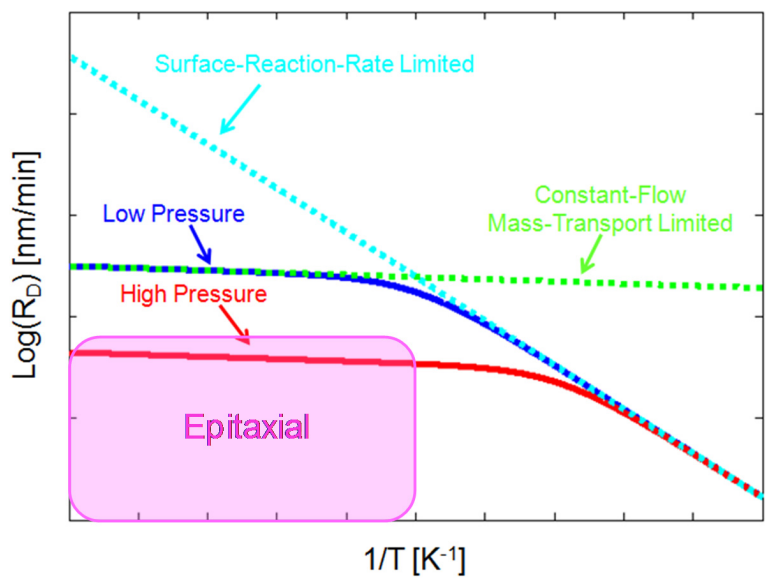

**Figure S2.** Representative plot of (constant-flow) CVD deposition rate, $R_D$, vs. inverse temperature, $1/T$.

In order for CVD films to grow epitaxially, the surface diffusion length, $L_D \propto (R_D)^{-\frac{1}{2}}\exp(-E_A/k_B T)$, of a given species must be maximized. This can be achieved by decreasing the deposition rate, $R_D$, and/or increasing $T$. In general, $R_D = k_s h_g (k_s + h_g)^{-1} \times (C_G/N)$, where $C_G$ is the gas-phase concentration of reactive species, $N$ is the number of species per unit volume of the deposited film ($1.9 \times 10^{22}$ cm$^{-3}$ for MoS$_2$), $h_g \propto (P_{tot})^{-1}$ is the mass transfer coefficient (inversely proportional to total system pressure), and $k_s \propto \exp(-E_A/k_B T)$ is the surface reaction rate (very strongly dependent on $T$) [1].

From the equation for $R_D$, it is evident that, for given values of $C_G$ and $N$, $R_D$ will be limited by either $k_s$ or $h_g$. For best process control, it is easiest to operate in a mass-transport limited regime, and to this end increasing $T$ is desirable. Additionally, increasing the total ambient pressure of the system greatly decreases $R_D$ when growth is mass-transport limited, and so increased pressure (in this case, ~760 Torr) is also advantageous. The biggest difference between the ideal case shown in Figure S2 and the CVD processes employed in this work is the use of solid precursors, which causes $C_G$ to be a function of precursor mass, temperature, pressure, and even time. Thus the exact amount of solid precursors must be simultaneously tuned with growth conditions to achieve uniform 1L deposition across large areas. Finally, the increased pressure causes the mean free path of gaseous reactive species, $\lambda \propto (k_B T/P_{tot})$ to be greatly reduced. Thus the solid precursor must be evenly distributed directly underneath the growth substrate to mimic showerhead injectors used for APCVD (atmospheric pressure CVD) with gaseous precursors.

With these points in mind, a more optimized growth recipe that allows for continuous mostly monolayer (~90% 1L by area while maintaining large single crystal sizes, as in Fig. 1 of the main text) was established after the device fabrication. (Electrical data for these growths will be provided in a future manuscript.) The experimental setup for this recipe involved separating the precursors by ~10 inches, reducing the MoO$_3$ amount, and using hydrophobic SiO$_2$/Si treated with HMDS to ensure that PTAS droplets consume as little area as possible. 20 - 30 μL of 100 μM PTAS is placed in 2 - 5 μL droplets around the edges of the growth substrate (seen as "coffee ring" circles after growth, in Fig. 1a of the main text), and 0.2 - 0.8 mg of $MoO_3$ are used to tailor the growths to achieve large individual $MoS_2$ crystals or a continuous film, respectively. This recipe is:

1) Pump the tube to base pressure (~100 mTorr)

2) Flush the tube with 1500 sccm Ar and then close the butterfly valve to reach 760 Torr

3) Ramp to 400 °C in 10 minutes and reduce the Ar flow to 30 sccm*

4) Ramp to 850 °C in 20 minutes and hold for 15 minutes

5) Allow the furnace to cool to 650 °C before opening the hatch for rapid cooling

*note: 30 sccm in a 2-inch tube corresponds to 7.5 sccm in a 1-inch tube.

The PTAS seeding layer is essential for obtaining large-area $MoS_2$ growth on the $SiO_2$ surface. Similar to [2], only islands of small (<500 nm) $MoS_2$ particles could typically be grown on $SiO_2$ substrates at any temperature without the use of PTAS.

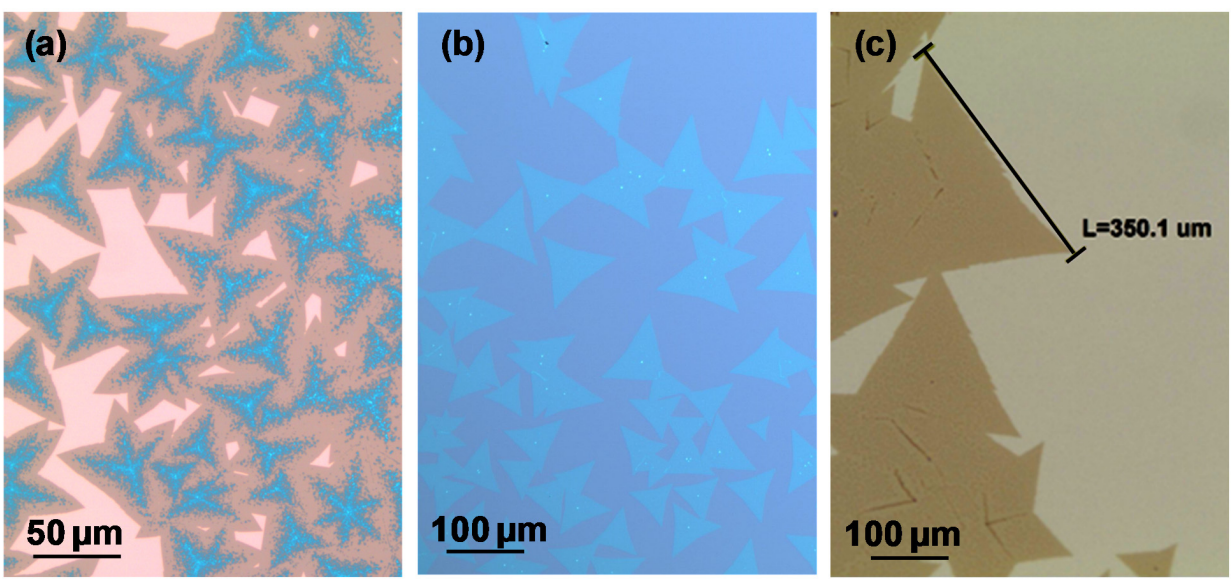

**Figure S3** Optical images of different $MoS_2$ growths. (a) The growth that yielded the devices discussed in the main text, grown on 90 nm $SiO_2$ on Si. (All devices were fabricated on the monolayer regions.) (b) The optimized growth that allows for very large, all monolayer $MoS_2$ triangles similar to Fig. 1a, but grown on 300 nm $SiO_2$. (c) Edge of a continuous $MoS_2$ film from a tailored growth showing 1L single-grain sizes in excess of 350 μm on an edge, grown on 30 nm $SiO_2$ on Si. Darker regions in continuous films are small bilayer regions that can appear at highly-misoriented grain boundaries or when the size of the individual crystals exceeds the surface diffusion length for the given growth conditions.

## B. Additional Electrical Data and Transfer Length Calculations

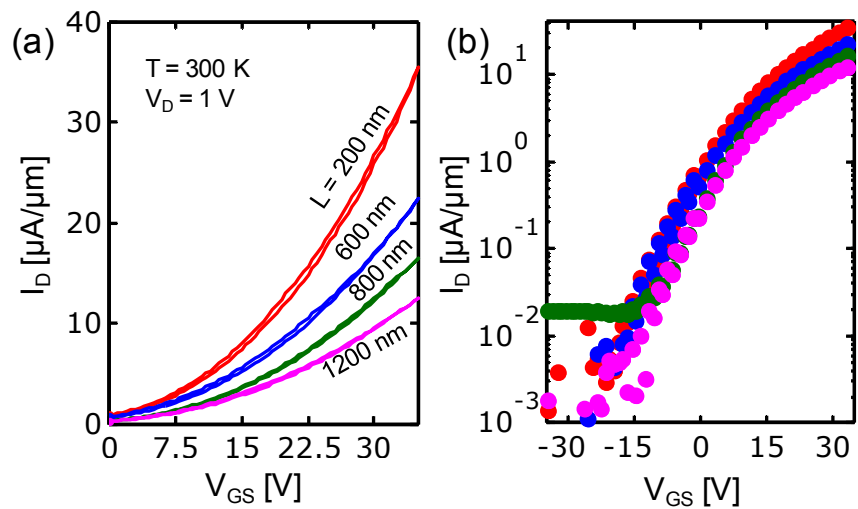

**Figure S4.** (a) Forward and backward $I_D$-$V_{GS}$ sweeps demonstrating very little hysteresis of our devices in vacuum. (b) Measured data for the same transfer curves shown in (a) plotted in log scale, showing $I_{ON}/I_{OFF} \geq 10^4$ (partly limited here by the measurement range of the Keithley 4200 semiconductor parameter analyzer).

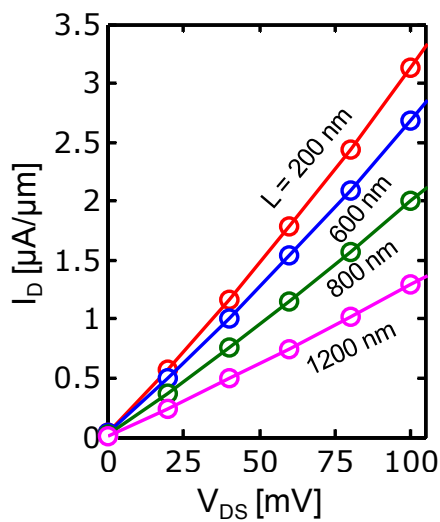

**Figure S5.** Low-field $I_D$-$V_{DS}$ sweeps are linear but do not necessarily indicate Ohmic contacts in terms of band alignment at the contacts. As extracted in the main text, there is still a non-negligible contact resistance due to a Schottky barrier at the Au-MoS$_2$ interface.

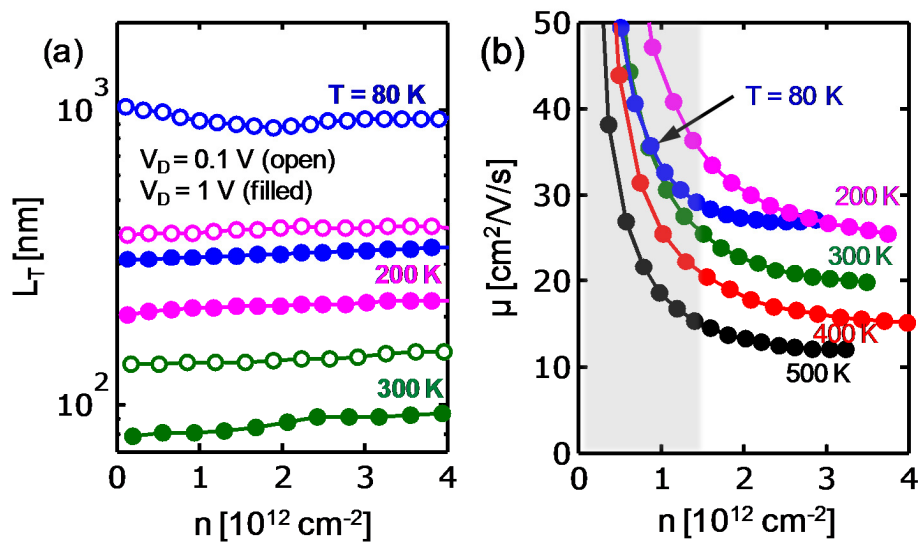

**Figure S6.** (a) Contact transfer length vs. carrier density in log scale. The trend is similar to that in Fig. 4b of the main text due to the dependence of $L_T$ on $\rho_C$ in Equation 3. (b) $\mu_{eff}$ vs. $n$ for various $T$. The gray shaded region indicates $\mu_{eff}$ with errors $\geq 30\%$ for low $n$ due to the assumed linear inversion charge model.

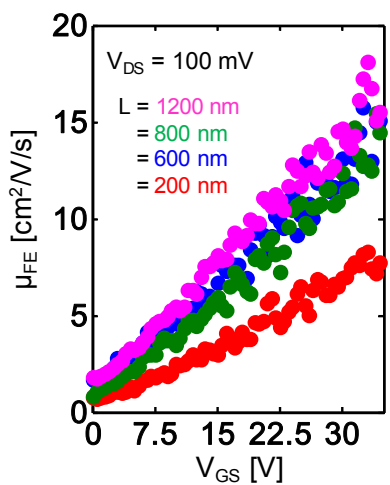

**Figure S7.** Field-effect mobility extractions for the same devices in Fig. 3a of the main text, here with $V_{DS} = 0.1$ V. Although the phenomenon of "peak mobility" is not exhibited in our devices, for all values of $V_{GS}$, $\mu_{FE} < \mu_{eff}$, similar to the simulations in Fig. S8b. Note that shorter channel devices exhibit lower $\mu_{FE}$ due to the larger contribution of $R_C$ to total device resistance.

## C. Simulations of Extracted Mobility Values

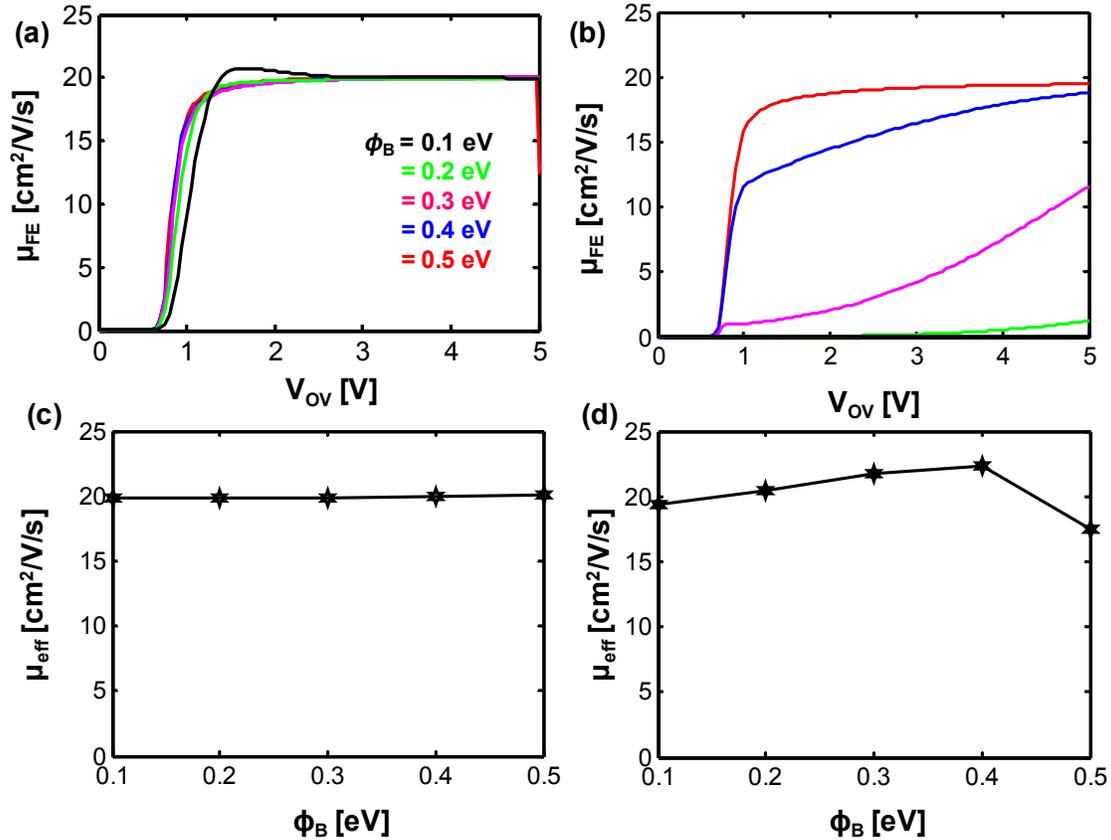

**Figure S8.** (a) and (b) Sentaurus simulated *field-effect mobility* extractions plotted against over-drive voltage ($V_{OV} = V_{GS} - V_T$) for varying Schottky barrier heights and tunneling masses of $0.01m_0$ and $1.0m_0$, respectively. $\mu_0 = 20$ cm²/V/s and $V_{DS} = 0.1$ V for all simulations. Both over- and under-estimation of the "true" mobility can be seen for various contact resistance parameters. (c) and (d) Sentaurus simulated *effective mobility* extractions from TLM structures plotted against varying Schottky barrier heights for tunneling masses of $0.01m_0$ and $1.0m_0$, respectively. $\mu_0 = 20$ cm²/V/s and $V_{DS} = 1.0$ V for all simulations. The "true" channel mobility is extracted for all combinations of $m_{tun}$ and $\phi_B$.

## D. Brief History of Electrical Results Reported for Synthetic Monolayer MoS₂

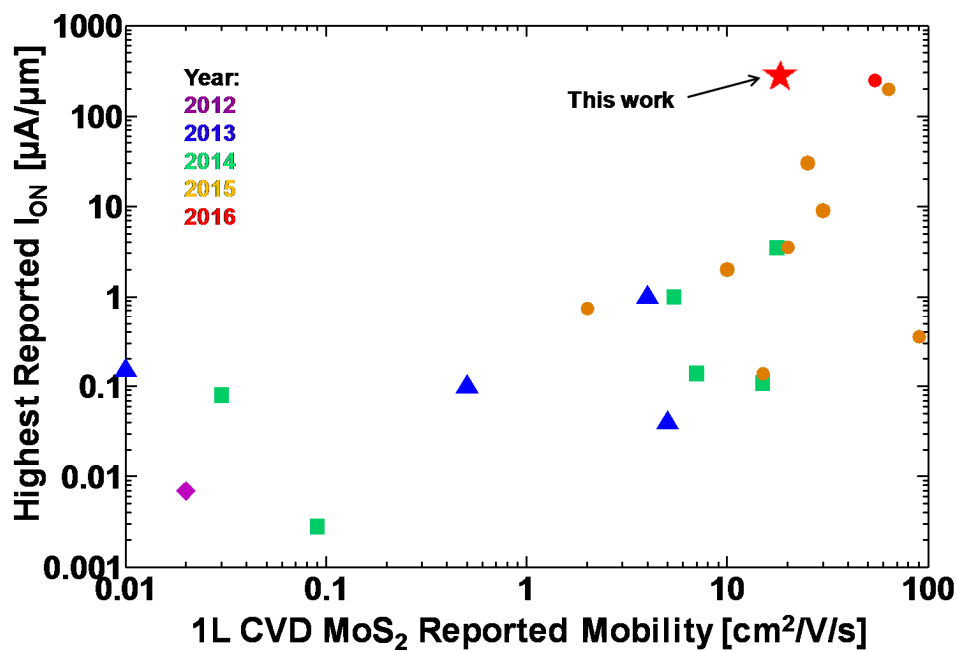

**Figure S9.** Historical plot of 1L CVD MoS₂ devices with highest reported drive currents and reported mobility values [2-22]. The type of mobility value reported varies by source, and this plot includes two-terminal and four-terminal field-effect mobility, two-terminal field-effect mobility with contact resistance estimated and subtracted, Y-function method [23], effective mobility extracted from TLM devices (this work, generally assumed to be more reliable), and mobility fit to velocity saturation models. References for these data are given in the Supplement References section below, in order of ascending mobility.

## E. Estimation of the mean free path ($\lambda_{MF}$) in MoS₂

We can estimate the mean free path as $\lambda_{MF} = v_{2D}\tau_C$, where $v_{2D} = (\pi k_B T/2m_{eff})^{1/2}$ is the average thermal velocity of electrons and $\tau_C = \mu_{eff}m_{eff}/q$ is the average collision time for carriers. Here, $k_B$ is the Boltzmann constant, $T$ is the average device temperature, $\mu_{eff}$ is the effective mobility at the temperature $T$, $q$ is the elementary charge, and $m_{eff}$ is the in-plane carrier effective mass. For electrons in monolayer MoS₂, $m_{eff} = 0.48m_0$ where $m_0$ is the mass of free electron.

As can be seen from Fig. S10, the mean free path for electrons is well below the channel lengths of the MoS₂ studied in this paper ($\lambda_{MF} \sim 1$ to 3 nm for $\mu_{eff}$ of 20 to 80 cm²/V/s). The low mean free path for electrons, therefore, justifies the use of semi-classical transport for the model used for MoS₂ performance projections.

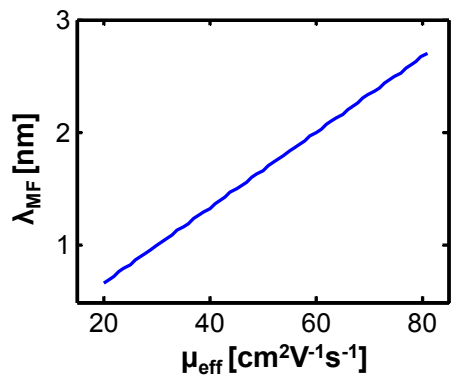

**Figure S10:** Estimated electron mean free path ($\lambda_{MF}$) MoS₂ vs. electron mobility ($\mu_{eff}$).

## F. Carrier confinement in 2D and 3D MOSFETs

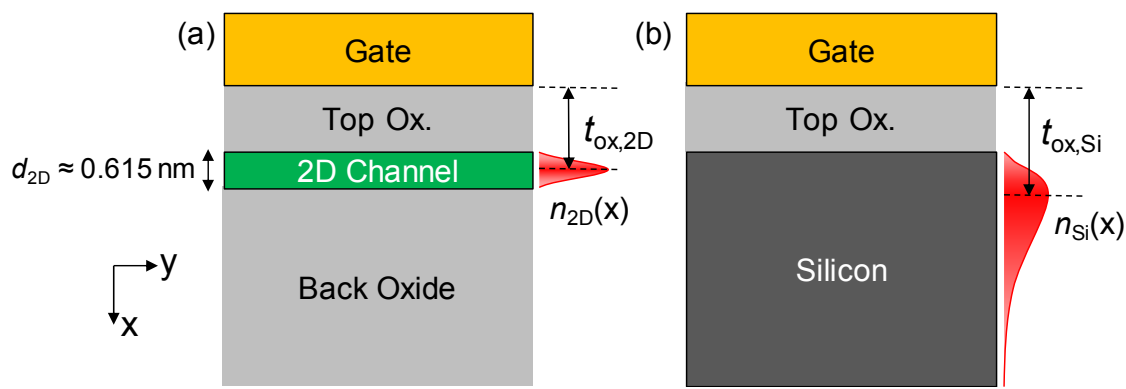

**Figure S11:** Schematic showing the effect of quantum confinement in MOSFET channel with: (a) monolayer MoS₂ and (b) traditional bulk materials like silicon. The red curve shows the carrier density as a function of the channel depth ($x$). In 2D monolayer channels, the effective oxide thickness $t_{ox,2D} < t_{ox,Si}$ for similar oxide physical thickness and gate voltage overdrive.